\documentclass[reprint,showkeys,showpacs,amsmath,amssymb,aps,prd,nofootinbib]{revtex4-1}
\usepackage{color}
\usepackage[pdftex]{hyperref}
\begin{document}
\title[Cosmological magnetic field]{Cosmological magnetic field---the boost-symmetric case}
\author{Ji\v{r}\'{i} Vesel\'{y}}%
 \email{jiri.vesely@utf.mff.cuni.cz}
\author{Martin \v{Z}ofka}
\email{zofka@mbox.troja.mff.cuni.cz}
\affiliation{Institute of Theoretical Physics, Faculty of Mathematics and Physics, Charles University, Czech Republic}
\vspace{10pt}
\begin{abstract}
We find a class of cylindrically symmetric, static electrovacuum spacetimes generated by a non-homogeneous magnetic field and involving the cosmological constant and one additional parameter, which determine uniquely the strength of the magnetic field. We provide a simple model of a source producing the field.
\end{abstract}
\pacs{04.20.Jb, 04.40.Nr}
\keywords{Einstein--Maxwell equations, cylindrical symmetry, magnetic field, cosmological constant}
\maketitle
\section{\textbf{Introduction}}
With the notable exception of black holes, Einstein equations of general relativity are notoriously difficult to solve for exact solutions that would describe an astrophysically relevant situation with all its nuanced details.
Therefore, one usually reduces the problem at hand assuming various symmetries and solves the resulting, simpler equations. If this approach does not work one needs to resort to numerical relativity, always making sure the results do correspond to our expectations based on exact solutions again, which are thus important both as testbeds for approximations and numerical calculations and as a rough model of highly relativistic observable objects. The simplest approach, of course, is to deal with the gravitational field only and assume a vacuum solution but the next step is to include, for instance, a gravitating fluid or electromagnetic field in a self-consistent manner, taking into account its back-reaction on the gravitational field. The inclusion of an electromagnetic field is important not only from the mathematical perspective but also because of its astrophysical relevance due to its role in the physics of fields and particles in the vicinity of compact objects requiring a general relativistic description \cite{Petri,Carrasco+Palenzuela+Reula} and also in view of the magnetic fields observed to permeate the intergalactic space \cite{Tavecchio,Neronov-Vovk}, which are presumably of primordial origin \cite{Subramanian}.

This paper investigates a spacetime involving an electromagnetic field, assuming further that the solution of the relevant Einstein--Maxwell equations is static and cylindrically symmetric. We presume that the electromagnetic field inherits the symmetry of the spacetime and that it is aligned with the axis of symmetry, following thus in the footsteps of the Bonnor--Melvin magnetic solution \cite{Bonnor,Melvin} where the magnetic field strength varies with position---indeed, since the field gravitates, a constant field throughout the spacetime would necessarily collapse onto itself. The Bonnor-Melvin solution has been of renewed interest recently as it is often used as a non-spherical background for studies of black holes immersed in a magnetic field \cite{Astorino+Compere+Oliveri+Vandevoorde,Brito+Cardoso+Pani}, its analogies are explored in generalized theories of gravity \cite{Bambi+Olmo+Rubiera-Garcia}, cylindrical counterparts of (anti-)photon spheres are investigated in the spacetime \cite{Gibbons+Warnick}, and it even serves as a seed that generates solutions interpolating between early- and late-time anisotropic cosmological models \cite{Kastor+Traschen}.

We have recently studied a solution of Einstein--Maxwell equations that generalizes the Bonnor--Melvin universe to the case of a non-zero cosmological constant $\Lambda$ \cite{Zofka}, which counters the gravitational pull of the magnetic field, enabling the resulting balance of the static solution while keeping the magnetic field's invariant constant everywhere. Therefore, the solution is the best general relativistic analog of a classical constant magnetic field.
It has the form of a direct product of a two-dimensional Minkowski spacetime and a 2-sphere of constant radius $1/\sqrt{2\Lambda}$. This is typical for compactified spacetimes resulting from low-energy approach to higher-dimensional solutions due to the string theory \cite{Prasetyo2015, Emparan}. It is fitting in this respect that the magnetic field has the form of the Dirac monopole \cite{Milton_2006}. It would be of interest to put this family of spacetimes in a broader perspective as a member of a wider class of solutions with a clear physical meaning
and it is thus natural to ask whether there is a more general solution with the same symmetries but with a varying magnetic field that would include as special cases both the homogeneous-magnetic-field spacetime and the Bonnor--Melvin solution with $\Lambda = 0$. We present the solution here
and also provide a simple physical model generating the field: a cylindrical shell forming a massive and current-carrying relativistic solenoid running along the axis of symmetry. In fact, one would expect the spacetime to very roughly approximate the situation in the vicinity of any current carrying body that is locally approximately cylindrically symmetric such as a slowly rotating charged sphere along its equator.



The paper is organized as follows: in Section \ref{General Einstein--Maxwell equations}, we establish the coordinate system and the form of the metric and electromagnetic field tensors and present the Einstein--Maxwell equations. In Section \ref{General solution}, we then reduce the set of equations to a single 3\textsuperscript{rd}-order differential equation and discuss the number of independent parameters of the solution. Section \ref{Symmetric case} then focuses on a subfamily of solutions and further simplifies the single remaining equation to the 2\textsuperscript{nd} order. We continue with Section \ref{Exact solution} where we present the most general solution of the resulting equation, specifying the metric and discussing the geometry and physics of the solution. We explain its relation to previously obtained exact solutions featuring the same symmetry of the gravitational and electromagnetic fields. Section \ref{Shell-source model} investigates a simple shell-source model producing the discussed fields and yielding an interpretation of the solution in terms of streams of charged, massive particles. We briefly summarize our results in Section \ref{Conclusions}.
\section{\textbf{Einstein--Maxwell equations}}\label{General Einstein--Maxwell equations}
The metric of a general static, cylindrically symmetric spacetime can be written as
\begin{equation}\label{general line element}
\mathrm{d}s^2 = -\exp A(r) \; \mathrm{d}t^2 + \mathrm{d}r^2 + \exp B(r) \; \mathrm{d}z^2 + \exp C(r) \;\mathrm{d}\varphi^2,
\end{equation}
where $r \in \! I\hspace{-0.13cm}R^+$ is the proper radial distance, $t,z \in \! I\hspace{-0.13cm}R$ are temporal and azimuthal coordinates, and $\varphi \in [0,2\pi)$ measures the angle around the axis of symmetry. We are looking for a self-consistent solution generated by a magnetic field aligned with the axis of symmetry
\begin{equation}\label{Maxwell original}
    F = H(r) \: \mbox{d}r \wedge \mbox{d}\varphi,
\end{equation}
yielding
\begin{equation}\label{Maxwell invariant}
    F_{\mu\nu}F^{\mu\nu} = 2H^2\mbox{e}^{-C} \equiv 2f^2,
\end{equation}
where we defined a new quantity, $f(r)$, while $\star F_{\mu \nu} F^{\mu \nu} = 0$. For details, we refer the reader to our previous paper \cite{Zofka} while here we just briefly recall the Einstein equations
\begin{eqnarray}
  2(B'' + C'') + \left( B' \right)^{2} + \left( C' \right) ^{2} + B'C' + 4\Lambda + 4f^2 &=& 0,\;\;\;\;\;\label{Einstein t}\\
  2(A'' + C'') + \left( A' \right)^{2} + \left( C' \right) ^{2} + A'C' + 4\Lambda + 4f^2 &=& 0,\label{Einstein z}\\
  2(A'' + B'') + \left( A' \right)^{2} + \left( B' \right) ^{2} + A'B' + 4\Lambda - 4f^2 &=& 0,\label{Einstein phi}\\
  A'B' + A'C' + B'C' + 4\Lambda - 4f^2 &=& 0.\label{Einstein r}
\end{eqnarray}
with primes denoting derivative with respect to the radial coordinate. The source-free Maxwell equations are mostly satisfied identically while $ \sqrt{-g} {F^{\varphi \alpha}}_{;\alpha} = 0$ is a consequence of Einstein equations and it yields
\begin{equation}\label{Maxwell constraint}
    \mbox{e}^{\frac{A+B}{2}} f = const.
\end{equation}

What is the mathematical structure of the Einstein equations (\ref{Einstein t})--(\ref{Einstein r})? By adding (\ref{Einstein r}) to (\ref{Einstein t}) and to (\ref{Einstein z}) and by subtracting it from (\ref{Einstein phi}), we obtain a coupled system of 3 second-order ordinary differential equations for $A, B,$ and $C$ (in fact, it is a first-order system for $A', B',$ and $C'$).
\begin{eqnarray}
  2(B + C)'' + A'(B + C)'+\left[ (B + C)' \right] ^{2} + 8\Lambda &=& 0,\;\;\;\;\;\label{Einstein t no f}\\
  2(A + C)'' + B'(A + C)'+\left[ (A + C)' \right] ^{2} + 8\Lambda &=& 0,\;\;\;\;\;\label{Einstein z no f}\\
  2(A + B)'' + \left( A' \right)^{2} + \left( B' \right) ^{2} - C'(A + B)' &=& 0,\label{Einstein phi no f}
\end{eqnarray}
We can rescale the time and axial coordinates to make the values of both $A$ and $B$ vanish at an arbitrary radius. The value of $C$ determines the proper length of orbits of the Killing vector $\partial/\partial\varphi$. We thus have three integration constants corresponding to the first derivatives of the sought functions at a given initial radius. These are constrained physically through (\ref{Einstein r}), which determines $f^2$---the strength of the magnetic field at that location. We then solve the system of equations (\ref{Einstein t no f})--(\ref{Einstein phi no f}). If it so happens that the resulting spacetime includes an axis where $C \rightarrow -\infty$, see below, then we also require elementary flatness at the axis, yielding another constraint on the initial values. Therefore, the solution is completely determined by the cosmological constant, one free parameter, and two additional factors: geometry---the length of the hoop---and physics---the energy density of the magnetic field at a selected location.
\section{General solution}\label{General solution}
It turns out that the system (\ref{Einstein t})--(\ref{Einstein r}) can be separated in the following way. First, add (\ref{Einstein   phi}) and (\ref{Einstein r}) and substitute for $(A+B)'$ from (\ref{Maxwell constraint}) to yield
\begin{equation}
    -2 \frac{f''}{f} + 4 \frac{(f')^2}{f^2} + 4(\Lambda-f^2) - \frac{C'f'}{f} = 0.
\end{equation}
From here, we can express $C'$ as follows
\begin{equation}\label{formula_for_C'}
    C' = -2 \frac{f''}{f'} + 4 \frac{f'}{f} + 4\frac{f}{f'}(\Lambda-f^2).
\end{equation}
We now combine (\ref{Einstein   t}) + (\ref{Einstein   z}) $-$ (\ref{Einstein   phi}) $-$ (\ref{Einstein   r}) to find
\begin{equation}
    2C'' + (C')^2 - A'B' +8f^2=0,
\end{equation}
and express $A'B'$ from the resulting equation to substitute it together with (\ref{formula_for_C'}) and (\ref{Maxwell constraint}) into (\ref{Einstein r}) to obtain a single, separate, third-order equation for $f$
\begin{widetext}
\begin{equation}\label{full equation for f}
     f'''f' - 2(f'')^2 + f''\left(6f(\Lambda-f^2) + \frac{(f')^2}{f}\right) + (f')^2(11f^2 - 9\Lambda) - 4f^2(\Lambda-f^2)^2=0.
\end{equation}
\end{widetext}
We solve this equation and insert the solution into (\ref{formula_for_C'}). We now know both $f$ and $C'$. Writing (\ref{Einstein r}) as $A'B' + (A + B)'C' + 4(\Lambda - f^2) = 0$, we substitute here for $(A+B)'$ from (\ref{Maxwell constraint}), which also yields $B'=-A'-2f'/f$. We finally have a first-order equation for $A$, which is quadratic in $A'$, and the same equation for $B$ if we substitute for $A$ instead. Its solution reads
\begin{eqnarray}
  A' &=& -\frac{f'}{f} \pm \sqrt{4\frac{f''}{f}-7\left(\frac{f'}{f}\right)^2-4(\Lambda-f^2)},\label{differential equation for A}\\
  B' &=& -\frac{f'}{f} \mp \sqrt{4\frac{f''}{f}-7\left(\frac{f'}{f}\right)^2-4(\Lambda-f^2)}.\label{differential equation for B}
\end{eqnarray}
Again, we determine $f$ from (\ref{full equation for f}), requiring 3 initial conditions, and then we calculate $A,B,$ and $C$ from (\ref{differential equation for A}), (\ref{differential   equation for B}), and (\ref{formula_for_C'}), where we use the rescaling of $t$ and $z$ and the hoop length, so there are no additional integration constants. As mentioned in the previous section, we further require a regular axis and a particular energy density of the magnetic field at a given location, leaving us with one free integration constant. Therefore, apart from the cosmological constant, the hoop length and field strength at a chosen point and one additional constant are the independent parameters of the solution.


%
\section{Symmetric case}\label{Symmetric case}
To further reduce the master equation (\ref{full equation for f}), we assume now $A=B$. This corresponds to a spacetime which is a warped product of a conformal 2D Minkowski and an additional 2D space. Equations (\ref{Einstein   t}) and (\ref{Einstein   z}) then coincide. The Maxwell equation (\ref{Maxwell constraint}) yields
\begin{equation}\label{A as a function of f}
    A = \mbox{const} - \ln f.
\end{equation}
The remaining Einstein equations read
\begin{eqnarray}
  2(A'' + C'') + \left( A' \right)^{2} + \left( C' \right) ^{2} + A'C' + 4\Lambda + 4f^2 &=& 0,\;\;\;\;\;\;\;\label{Einstein_z_A=B}\\
  4A'' + 3\left( A' \right)^{2} + 4\Lambda - 4f^2 &=& 0,\label{Einstein_phi_A=B}\\
  \left( A' \right)^{2} + 2A'C' + 4\Lambda - 4f^2 &=& 0.\label{Einstein_r_A=B}
\end{eqnarray}
Taking the difference (\ref{Einstein_phi_A=B}) $-$ (\ref{Einstein_r_A=B}), we have
\begin{equation}
    2A'' + \left( A' \right)^{2} - A'C'=0,
\end{equation}
which can be integrated to yield
\begin{equation}\label{C as a function of A}
    C = \mbox{const} + A + 2 \ln A'.
\end{equation}
We are left with two Einstein equations but if we insert the above expressions for $C$ and $A$ in terms of $f$, the equations are not independent and we finally have a single second-order equation for $f$
\begin{equation}\label{equation_for_f}
    4ff'' -7\left( f' \right)^{2} -4f^2(\Lambda - f^2)=0.
\end{equation}
Plugging (\ref{A as a function of f}) into (\ref{C as a function of A}), we find
\begin{eqnarray}
  \exp A(r) &=& \exp B(r) = \frac{\alpha}{f(r)},\label{A and B in terms of f}\\
  \exp C(r) &=& \alpha \beta \frac{f'(r)^2}{f(r)^3},\label{C in terms of f}
\end{eqnarray}
with $\alpha, \beta$ integration constants. The last relation implies that $f'(r)=0$ defines an axis. We can always redefine the radial coordinate by shifting it arbitrarily and thus we can put the axis at $r=0$.

\section{Exact solution}\label{Exact solution}
Let us now rewrite (\ref{equation_for_f}) as follows
\begin{equation}\label{equation_for_f''}
    f'' =\frac{7}{4}\frac{\left( f' \right)^{2}}{f} + f(\Lambda - f^2)
\end{equation}
Equation (\ref{equation_for_f''}) lacks $r$ and we thus use $f$ as the independent variable. Let us define $v=f'$, which yields $f'' = (f')' = \mathrm{d}v/\mathrm{d}r = (\mathrm{d}v/\mathrm{d}f)(\mathrm{d}f/\mathrm{d}r) = (\mathrm{d}v/\mathrm{d}f)v$ and, together with (\ref{equation_for_f''}), we get
\begin{equation}\label{first-order equation}
  v(f)\frac{\mathrm{d}v(f)}{\mathrm{d}f} = \frac{7}{4}\frac{v(f)^{2}}{f} + f(\Lambda - f^2).
\end{equation}
We first reduce the last equation to a linear equation through the substitution $v^2=w$ and thus $\mathrm{d}w/\mathrm{d}f = 2v (\mathrm{d}v/\mathrm{d}f)$ to obtain
\begin{equation}\label{first-order linear}
  \frac{1}{2}\frac{\mathrm{d}w}{\mathrm{d}f} = \frac{7}{4}\frac{w}{f} + f(\Lambda - f^2).
\end{equation}
We now solve the homogeneous equation
\begin{equation}\label{homogeneous first-order equation}
  \frac{1}{w}\frac{\mathrm{d}w}{\mathrm{d}f} - \frac{7}{2}\frac{1}{f} =0
\end{equation}
to find $w_0=f^{7/2}$, which we use as an integration factor, dividing (\ref{first-order linear}) with it to produce
\begin{equation}\label{iuhoui}
  \frac{\mathrm{d}w}{\mathrm{d}f}f^{-\frac{7}{2}}-\frac{7}{2}wf^{-\frac{9}{2}} = \frac{\mathrm{d}}{\mathrm{d}f}\left(wf^{-\frac{7}{2}}\right) = 2 f^{-\frac{5}{2}}(\Lambda-f^2).
\end{equation}
This can be integrated to
\begin{equation}
  wf^{-\frac{7}{2}} = \gamma - 4 \sqrt{f} - \frac{4}{3}\Lambda f^{-\frac{3}{2}}
\end{equation}
and
\begin{equation}\label{the square root}
  v = \frac{\mathrm{d}f}{\mathrm{d}r} = \pm \sqrt{\gamma f^{\frac{7}{2}} - 4 f^4 - \frac{4}{3}\Lambda f^2},
\end{equation}
which can be separated again for us to finally write
\begin{equation}\label{inverse of f}
  r = \pm \int \frac{\mathrm{d}f}{\sqrt{\gamma f^{\frac{7}{2}} - 4 f^4 - \frac{4}{3}\Lambda f^2}},
\end{equation}
yielding $r=r(f)$, the inverse of the sought function. Using (\ref{C in   terms of f}) again, we find that the roots of the square root in (\ref{the square root}) where $f'$ vanishes determine the location of axes, but the integral (\ref{inverse of f}) exists precisely between two subsequent roots $f'=0$ where the argument of the square root is positive. Therefore, we always have two axes unless $f$ vanishes there.

Let us now change the radial coordinate and use the density of the magnetic field $f$ instead of $r$ via (\ref{the square root}), using relations (\ref{A and B in terms of f}) and (\ref{C in terms of f}) for the metric functions. In the following text, we use primes to denote derivatives with respect to the new radial coordinate, $x'(f) = \mathrm{d}x(f)/\mathrm{d}f$. After rescaling $t$ and $z$ and redefining $\alpha\beta \rightarrow \beta$, the transformed metric reads
\begin{widetext}
\begin{equation}\label{line element new coordinates}
\mathrm{d}s^2 = \frac{1}{f} \left(-\mathrm{d}t^2 + \mathrm{d}z^2 \right) + \frac{\mathrm{d}f^2}{\gamma f^{\frac{7}{2}} - 4 f^4 - \frac{4}{3}\Lambda f^2} + \beta\frac{\gamma f^{\frac{7}{2}} - 4 f^4 - \frac{4}{3}\Lambda f^2}{f^3} \mathrm{d}\varphi^2.
\end{equation}
\end{widetext}
There are two free constants in the solution, $\beta$ and $\gamma$, and the cosmological constant, $\Lambda$. The electromagnetic field is
\begin{equation}\label{Maxwell}
    F = \sqrt{\frac{\beta}{f}} \: \mbox{d}f \wedge \mbox{d}\varphi, \;\;\; A = 2\sqrt{\beta f} \: \mbox{d}\varphi.
\end{equation}
The Kretschmann scalar
\begin{equation}\label{Kretschmann}
  K = 56 f^4 - 12 \gamma f^\frac{7}{2} + \frac{3}{4} \gamma^2 f^3 + \frac{8}{3} \Lambda^2
\end{equation}
is bounded for a finite $f$ (see below) so there is no curvature singularity anywhere throughout the spacetime. Another interesting fact about the solution is that it admits both signs of the cosmological constant. The metric obviously requires the master function, $\mathfrak{M} \equiv \gamma f^{\frac{7}{2}} - 4 f^4 - \frac{4}{3}\Lambda f^2$, and $\beta f$ to be positive to retain its +2 signature. For $\Lambda>0$, this implies $\gamma>(16/3)\Lambda^{1/4}$ (then the single maximum of $\mathfrak{M}$ is positive), ensuring there is a single, finite interval of $f>0$ where $\mathfrak{M}>0$. The special value $\gamma = (16/3)\Lambda^{1/4}$ is discussed at the end of this section. The two corresponding roots $\mathfrak{M}(f_1)=\mathfrak{M}(f_2)=0$ are simple\footnote{The first derivative of the non-polynomial function $\mathfrak{M}$ does not vanish there. In fact, using $\mathfrak{M}(f_i)=0$, we find $\mathfrak{M}'(f_i)=2 f_i (\Lambda - f_i^2)$.} and they define the position of the axes since $g_{\varphi\varphi}=0$ there. The proper radial distance between the axes is finite and the spacetime is thus radially compact. In fact, it still corresponds to a product of a warped Minkowski and a compact 2D space. For $\Lambda<0$, any $\gamma$ is fine, yielding a single finite interval $f \in [0;f_0]$ with $\mathfrak{M}(0) = \mathfrak{M}(f_0)=0$, ensuring $\mathfrak{M} \ge 0$. The upper root is simple and represents an axis again in complete analogy with the $\Lambda>0$ case above, while at $f=0$ we have $\mathfrak{M} = \mathfrak{M}' = 0$ and this is in fact the asymptotic region since $g_{\varphi\varphi}$ diverges here and its proper distance from any other $f$ is infinite. What is the asymptotic form of the spacetime as $f\rightarrow 0$? To the lowest order, we obtain
\begin{equation}
\mathrm{d}s^2 = \frac{1}{f} \left[ \left(-\mathrm{d}t^2 + \mathrm{d}z^2 \right) + \frac{\mathrm{d}f^2}{\frac{4}{3}|\Lambda| f} + \beta\frac{4}{3}|\Lambda| \mathrm{d}\varphi^2 \right].
\end{equation}
Redefining the radial coordinate, we obtain
\begin{equation}
\mathrm{d}s^2 = \frac{1}{\rho^2} \left[ -\mathrm{d}t^2 + \mathrm{d}z^2 + \mathrm{d}\rho^2 + \mathrm{d}\eta^2 \right],
\end{equation}
which is the anti-de Sitter spacetime. One can also consider negative values of $f$ but the situation is identical with $\beta$  and (a purely imaginary) $\gamma$ simply changing their signs and $t$ and $z$ switching their meanings.

Summarizing the geodetic structure of the spacetime, we conclude there is a circular null geodesic located at $f=\left(\frac{3}{16}\gamma \right)^2$. Axial null geodesics exist everywhere and for any parameters of the spacetime. A radial null ray has a vanishing coordinate velocity at any axis regardless of the sign of the cosmological constant and it takes a finite affine parameter to reach the axis\footnote{In the vicinity of an axis $f=f_i>0$ with $\mathfrak{M}(f_i)=0$, we find $(f-f_i) \approx (\eta - \eta_0)^2$ with $\eta$ the affine parameter.}. For $\Lambda<0$, the coordinate velocity of a radial null ray vanishes at $f=0$ and it takes an infinite affine parameter but a finite coordinate time to reach $f=0$ where the magnetic field vanishes.

The requirement of elementary flatness near an axis located at $f=f_i > 0$ where $g_{\varphi\varphi}=0$ reads
\begin{equation}\label{elementary flatness}
  \sqrt{g_{\varphi\varphi}} \approx \int_{f_0} \sqrt{g_{ff}} \; \mathrm{d}f.
\end{equation}
Expanding $\mathfrak{M}(f)$ near its root at $f = f_i$ and using the fact that $\mathfrak{M}(f_i)=0$ to express $\mathfrak{M}'(f_i)$, we can write for the integral
\begin{equation}
  \int_{f_i}^f \frac{\mathrm{d}x}{\sqrt{\mathfrak{M}'(f_i)(x-f_i)}} = 2 \sqrt{\frac{f-f_i}{\mathfrak{M}'(f_i)}} = 2 \sqrt{\frac{f-f_i}{2 f_i (\Lambda - f_i^2)}} \;,
\end{equation}
while the left-hand side of (\ref{elementary flatness}) gives
\begin{equation}
  \sqrt{\frac{\beta}{f_i^3}\mathfrak{M}'(f_i)(f-f_i)} = \sqrt{\frac{2\beta}{f_i^2}(\Lambda-f_i^2)(f-f_i)}\;.
\end{equation}
Relation (\ref{elementary flatness}) thus fixes one integration constant as
\begin{equation}\label{beta fixed simplified}
  \beta = \frac{f_i}{(f_i^2-\Lambda)^2}.
\end{equation}
For a positive cosmological constant, we would like to comply with the above equation for the two separate axes. Remember that $f_i=f_i(\Lambda,\gamma)$ so that (\ref{beta fixed simplified}) evaluated at both axes would produce two equations determining $\gamma$ as well as $\beta$ and the entire solution would be given in terms of $\Lambda$ only. This, however, is not possible since the right-hand sides of the two copies of (\ref{beta fixed simplified}) are not independent for the two axes and, in fact, we cannot have both axes regular at the same time---one involves a conical defect. For a negative cosmological constant, we only have one axis to deal with and, therefore, $\beta$ is determined in terms of $\gamma$, which remains a free parameter (or vice versa, of course), still ensuring a regular axis. Ultimately, the solution involves one free parameter in addition to the cosmological constant regardless of its sign.


The spacetime is type D everywhere apart from $f=0$ and $f=(\gamma/8)^2$, where it is type O. It is boost-rotation symmetric as expected since it is conformal to $M_2 \times I\hspace{-0.13cm}R_2$. The solution belongs to the Kundt class. Specifically, it can be brought into the form of a Pleba\'{n}ski--Demia\'{n}ski metric with two non-expanding repeated principal null congruences, which is then a Kundt spacetime of type D. Our metric (\ref{line element new coordinates}) can be transformed into the form (16.27), p. 316 of \cite{Griffiths+Podolsky}, by the transformation $f=1/p^2$. In the language of \cite{Griffiths+Podolsky}, we have $\gamma=0, \rho=p, \epsilon_2=0, \epsilon_0=1, \epsilon=0, m=0,$ and $k=-e^2-g^2$. The master function written in terms of our parameters reads $\mathcal{P}=-1 +(\gamma/4) p -(\Lambda/3)p^4$ while the 4-potential is $d\psi/p$ so that we conclude $e=0, g=-1, n=\gamma/8$, and $\alpha$ is arbitrary. To obtain the same spacetime, we need to unfold our angular coordinate $\varphi$ to cover the entire real axis and rescale it by $2\sqrt{\beta}$. Doing this, we lose the closed orbits of the angular Killing vector and thus also our original cylindrical symmetry. In the covering spacetime there is no need to restrict the value $\gamma$ through the requirement of elementary flatness. On the other hand, we still need to deal with the fact that $g_{\varphi\varphi}$ vanishes at two locations suggesting the cylindrical symmetry with two axes is a more natural interpretation. To our knowledge, this is the first member of this family apart from the Bonnor--Melvin \cite{Bonnor,Melvin} and Bonnor--Melvin-$\Lambda$ \cite{Zofka} solutions to have a clear physical meaning.

Let us look at two special cases: firstly, if $\Lambda=0$, we apply the transformation $f=2K(1-K^2 r^2)^2$, redefine $\gamma=4\sqrt{2K}$, and assume $\beta = 1/8K^3$ as required by the elementary flatness of the axis. This yields the metric
\begin{equation} \label{Bonnor Melvin}
g_{\mu \nu} = \alpha^{-2}(-\mbox{d}t^2+\mbox{d}z^2) + \alpha^{-5}\mbox{d}r^2 + \alpha r^2 \mbox{d}\varphi^2,
\end{equation}
with $t$ and $z$ rescaled, where
\begin{equation} \label{Definition of alpha}
\alpha = 1 - K^2 r^2
\end{equation}
and
\begin{equation} \label{Maxwell field invariant}
F_{\mu \nu} F^{\mu \nu} = 8 K^2 \alpha^4.
\end{equation}
This is the original Bonnor--Melvin solution as expected, see \cite{Griffiths+Podolsky}, p. 317. And, secondly, the value $\gamma=(16/3)\Lambda^{1/4}$ corresponds to a single value $f^2=\Lambda$ and yields the homogeneous solution of \cite{Zofka} where $f$ cannot be used as a coordinate since it has a constant value throughout the spacetime.

Note that an equivalent form of the metric has already been obtained as an example of usage of generalized Ernst's solution generating technique in \cite{Astorino}. Applying the transformation $t \rightarrow t \sqrt{\beta}$, $f \rightarrow \beta / ( 1 + \rho^2/4 )^2$ and $z \rightarrow z \sqrt{\beta}$ to our solution while keeping $\varphi$ the same, we obtain the metric (4.6) in \cite{Astorino} along with its corresponding electromagnetic potential, with $B=\beta$ and $k=\beta^{3/2} \gamma$.

\section{Shell-source model}\label{Shell-source model}
Let us address now the question of a physical source producing the field obtained above. We assume a cylindrically symmetric source that is both massive and charged. In fact, the simplest such a system is an infinitely thin cylindrical shell consisting of streams of oppositely charged particles moving along the surface of the shell according to the Israel junction conditions \cite{Israel, Kuchar} with a vanishing total charge but non-vanishing total electric current. Such a cylinder is the general relativistic analog of an infinite solenoid.

In classical physics, the solenoid has a homogeneous magnetic field inside while the field vanishes outside. Since the Maxwell field couples to the gravitational field in general relativity, the situation is different here. In fact, there is a solenoid solution where the spacetime (\ref{line element new coordinates}) forms the inside of the cylinder with a varying magnetic field aligned with a symmetry axis while outside we have the Linet--Tian spacetime \cite{Linet, Tian} with a vanishing magnetic field. This construction applies to both a positive and a negative cosmological constant. Note that we assume the cosmological constant to be the same on both sides of the shell and thus indeed constant throughout the spacetime although this is not a necessity in general. In both cases the induced stress-energy tensor $S_{ij}$ is diagonal in the Minkowski coordinates $(T, Z, \Phi)$ of the flat shell interface, with all entries positive for a range of $\Lambda$ and $\gamma$ and the induced 3-current only having a single non-zero entry, namely the azimuthal component. The temporal component of the stress-energy tensor dominates, $S_{TT} - S_{ZZ} - S_{\Phi\Phi}>0$, allowing an interpretation as due to four streams of massive, charged particles similarly to the discussion of the Bonnor--Melvin case with $\Lambda=0$ in \cite{Zofka+Langer}: we have two streams of positive particles spiralling up and down at the same rate along mirror-image paths ($v_Z \equiv {v_Z}_{(1)}=-{v_Z}_{(2)}, v_\Phi \equiv {v_\Phi}_{(1)}={v_\Phi}_{(2)}$) and two streams of negative particles following the complementary trajectories (${v_Z}_{(3)}=-{v_Z}_{(4)}=v_Z, {v_\Phi}_{(3)}={v_\Phi}_{(4)} = -v_\Phi$). Provided the charged dust streams are identical in terms of their rest-mass and charge densities, $\rho$ and $\sigma$, yielding a vanishing total electric charge, then their corresponding stress-energy tensor is also diagonal with non-negative entries and the 3-current $J_i$ only has the azimuthal component. Their densities and velocities thus represent 4 independent parameters of the model which can be fitted to the 3 diagonal entries of the induced energy momentum tensor and the single non-zero component of the induced 3-current
\begin{eqnarray}
  && \rho = \frac{1}{4}(S_{TT} - S_{ZZ} - S_{\Phi\Phi}),\\
  && \sigma =  \frac{J_\Phi}{4}\sqrt{\frac{S_{TT} - S_{ZZ} - S_{\Phi\Phi}}{S_{\Phi\Phi}}},\\
  && v_Z = \sqrt{\frac{S_{ZZ}}{S_{TT}}},  \;\;\;  v_\Phi = \sqrt{\frac{S_{\Phi\Phi}}{S_{TT}}}.
\end{eqnarray}
Consequently, the solution (\ref{line element new coordinates}) can be thought of as due to the counter-streaming charged massive particles spiralling along the solenoid and producing a static field. Here, we merely wished to provide a physically plausible source producing the field (\ref{line element new coordinates}) but the solenoid model can be discussed in further details. It would be of interest, for instance, to see how the model restricts the 2 free parameters of the spacetime. This, however, would mean an extensive numerical analysis while our focus in this paper is on the analytical approach so we leave this question open for now.

It is of interest that in \cite{Olea}, a similar approach was adopted in the case of 2+1 dimensional solutions investigated in \cite{Cataldo+Crisostomo+del Campo+Salgado}. Indeed, due to the translational symmetry along the axis of the spacetime discussed here, the Einstein--Maxwell equations effectively reduce to 2+1 dimensions and thus (\ref{line element new coordinates}) is related to the 2+1 solutions, which are also determined by the cosmological constant and one additional parameter like in 3+1 dimensions. However, a direct comparison is difficult since the 2+1 solutions only admit $\Lambda<0$ and the collapsing shell has a non-zero angular momentum as a result of the asymptotic properties of the outer spacetime. The magnetic field is non-vanishing both inside and outside of the shell and arises due to the shell's electric charge and rotation unlike in the solenoid model discussed here. The resulting equation of radial motion of the 1+1 shell can only be solved numerically and it is unclear whether the collapse would stop or bounce at a finite radius or whether it would continue to form some kind of a point singularity at the center. At any rate the static case needs to be dealt with separately along the lines presented above. A collapsing 2+1 shell model of the spacetime (\ref{line element new coordinates}) is certainly possible too but will rely on numerical calculations, which is beyond the scope of the discussion here.
\section{Conclusions}\label{Conclusions}
In this paper, we generalized our previous result on a constant magnetic field balanced by a positive cosmological constant to a spacetime involving a space-varying magnetic field determined by the cosmological constant and one additional parameter. The solution admits both signs of the cosmological constant and corresponds to a warped product of a 2D Minkowski and a 2D space. It also includes as special cases the Bonnor--Melvin solution and the constant field solution of \cite{Zofka}. We found a physically plausible source of the field in the form of an infinitely thin cylindrical shell consisting of streams of charged and massive particles spiralling along the surface of the shell. In the future we intend to study the general equations (\ref{full equation for f}), (\ref{differential equation for A}), and (\ref{differential equation for B}), dropping the requirement of boost symmetry in the $z$ direction. %
\begin{acknowledgments}
We thank Dr. Tom\'{a}\v{s} Ledvinka for fruitful discussions of the topic. We are grateful to Prof. Roberto Emparan for pointing out to us the higher-dimensional context of the homogeneous solution. J.V. was supported by Charles University, project GAUK 80918. M.Z. acknowledges funding by GACR 17-13525S.
\end{acknowledgments}

\end{document}